\documentclass[slac_one,10pt,twocolumn]{revtex4}
\usepackage{graphicx}
\usepackage{fancyhdr}
\usepackage[a4paper, top=1.3in, bottom=1.0in, left=0.6in, right=0.6in]{geometry}
\newcommand{\noi}{\noindent}

\newcommand{\be}{\begin{equation}}
\newcommand{\ee}{\end{equation}}
\newcommand{\D}{\Delta}

\begin{document}

\title{On the viability of SUSY GUTs with Yukawa unification:\\the crucial role of FCNCs} 
\author{Diego Guadagnoli}
\affiliation{Physik-Department, Technische Universit\"at M\"unchen,
D-85748 Garching, Germany}
\begin{abstract}
After a short overview of the hypothesis of Yukawa unification within SUSY GUTs, I detail on the general conditions for its phenomenological viability, 
emphasizing the decisive role played by observables in the flavor sector.
\end{abstract}

\preprint{TUM-HEP-701/08}

\maketitle

\thispagestyle{fancy}

\section{SUSY GUTs \& Yukawa Unification}

\noi The hypothesis of ``grand unification'' -- namely, that the strengths of the three Standard Model (SM) gauge couplings unify at some energy scale -- 
is not only appealing in itself, but allows also to address structural questions that in the SM remain unanswered, like the gauge quantum number 
assignments. This hypothesis is also supported by the measured low-energy values of the SM gauge couplings, if one supposes that above the Fermi 
scale $M_{\rm EW}$ the SM becomes supersymmetric: the predictions for the three couplings get closer at higher scales and eventually unify at a scale 
$M_G$ of about 3$\times 10^{16}$ GeV. Supersymmetry (SUSY) itself allows to keep the huge ratio $M_G/M_{\rm EW}$ natural beyond tree level.
The above ideas may be a red herring or else a major clue to the structure of fundamental interactions at scales that will probably never be 
tested at colliders. The second possibility looks favored by the (excellent) degree to which coupling unification works: this motivates the construction
of grand unified theories (GUTs) and the quest for further tests thereof.

The cornerstone predictions of SUSY GUTs are \cite{Raby-PDG}, besides {\em (i.)} gauge coupling unification at $M_G$, {\em (ii.)} the existence of SUSY 
throughout the energy interval between $M_G$ and $M_{\rm EW}$, and {\em (iii.)} proton decay. Concerning the latter, while a positive signal is 
generically expected to be behind the corner, a precise prediction is a highly model-dependent issue, depending in particular on whether the SUSY GUT 
under consideration is realized in 4 dimensions or via orbifold constructions \cite{proton-decay}.

Turning to {\em (ii.)}, the expectation of low-energy SUSY, the question of the predicted pattern of SUSY masses is, again, a model-dependent 
one, because of the unknown mechanism of SUSY breaking and of the form that Yukawa couplings assume at the high scale, in the first place. To make progress, 
one assumes universalities in soft SUSY-breaking terms at $M_G$, which can then be motivated in specific SUSY-breaking scenarios. A very elegant 
additional assumption, potentially testable in the SM {\em fermion} masses and mixings, is that of Yukawa unification (YU) at the 
GUT scale. It is motivated by the fact that, due to the higher degree of symmetry, matter fields must sit in appropriate representations of the gauge group, 
thereby sharing a common Yukawa coupling. Since this simple picture can be spoiled by e.g. the presence of higher-dimensional interactions, the crucial 
question is whether YU may leave any low-energy remnant at all. While for the light fermion generations this is definitely not the case, for the third 
generation it remains an open and appealing possibility. The latter immediately singles out SO(10) as the potentially most predictive case, in that it is 
the simplest group allowing to relate all third generation fermion masses to one another.

\section{Testing the viability of Yukawa Unification}

\noi The possibility of explaining the top-bottom mass ratio $m_t/m_b \gg 1$ with a unified Yukawa requires necessarily the vev's of the corresponding 
Higgs fields to satisfy $\tan \beta = v_U/v_D \gg 1$. In absence of a protective symmetry, however, such a hierarchy is subject to large radiative 
corrections, as elucidated in \cite{HRS}. In particular one can expect $m_b$ to receive corrections proportional to the `wrong' vev $v_U$. Since at 
tree level $m_b \propto v_D$, these corrections come with a factor of $\tan \beta$: they will be large exactly when one requires YU at tree level! It is 
then clear that, in the case of top-bottom YU, the $m_b$ prediction is intrinsically sensitive to the (unknown) SUSY spectrum and couplings, entering the 
mentioned radiative corrections. A workaround for this difficulty is to invert the strategy: instead of predicting quark mass relations from the theory 
parameters, use the measured masses and the requirement of YU to see whether a special region in the theory parameter space emerges \cite{BDR}. In fact 
it does. Under the simplifying assumption of universal GUT-scale soft terms for sfermions, $m_{16}$, and for gauginos, $m_{1/2}$, and with a 
{\em positive} $\mu$ parameter, YU prefers the region characterized 
by \cite{BDR,BF}\footnote{Quite interestingly the same relations emerge as fixed-point solution from the attempt to build SUSY models with radiatively-driven inverted 
scalar mass hierarchy (ISMH) \cite{BFPZ}, i.e. light third generation and heavy first and second generation sfermions. ISMH is an appealing possibility 
to relieve at one stroke the problem of fine tuning in the Higgs mass corrections, and of large flavor-changing neutral currents (FCNCs).} 
\be
-A_0\approx 2\,m_{16}, ~~ \mu,m_{1/2} \ll m_{16},
\label{ISMH}
\ee
where low-scale threshold corrections to $m_b$ are just the right $-$few\% needed. Note in fact that, in YU, $m_b$ corrections come dominantly 
from gluino ($\D m_b^{\tilde g}$ ) and chargino loops ($\D m_b^{\tilde \chi^+}$) and from a {\em non-decoupling}, O(+6\%) log term \cite{BDR}. 
Note as well that, individually, $\D m_b^{\tilde g}$ and $\D m_b^{\tilde \chi^+}$ are naturally O(40\%), and they cancel in half of the parameter space (that with 
$\mu > 0$). Hence it follows that, in viable parameter regions not characterized by squark decoupling, this cancellation {\em must} occur. Remarkably, the 
selected region is only the one in eq. (\ref{ISMH}) \cite{BDR}. Here a large $\D m_b^{\tilde \chi^{+}} < 0$ slightly overcomes the sum of the other, positive, 
contributions. The important insight \cite{BDR} that exact YU forces $m_b$ corrections to be non-zero, but way smaller than the generic expectation \cite{HRS}, 
actually holds independently from the requirements (\ref{ISMH}) \cite{TobeWells}.

How to further test this scenario? Note that, for exact YU, chargino contributions with sub-TeV SUSY masses are large also in the decay $b \to s \gamma$,
because they are proportional to $A_t \tan \beta$. Again, one would need cancellation patterns, considering the agreement between the $B \to X_s \gamma$ SM
prediction \cite{MisiakNNLO} and experiment \cite{HFAG}. Cancellations cannot occur with gluino contributions, which in the parameter space (\ref{ISMH}) are 
immaterial. One may envisage, instead, a cancellation with charged-Higgs contributions, that in fact is in principle possible in the same half of parameter space 
as chosen by $m_b$. However, also sensitive to heavy Higgs contributions is the branching ratio BR$(B_s \to \mu^+ \mu^-) \propto A_t^2 \tan^6 \beta / M_H^4$, via 
penguin diagrams mediated by heavy neutral Higgses \cite{Bsmumu-DP}. The stringent experimental bound \cite{Bsmumu-CDF} on this mode turns out to 
hinder the possibility of large enough charged-Higgs loops in $b \to s \gamma$.

The above discussion highlights the major role of the interplay $m_b$ -- FCNCs (especially $B$-decays) in deciding on the viability of YU. This question has
been addressed in many studies \cite{studiesYU}, with interesting differences in the approach and in the considered observables. 
One important difference is between bottom-up and top-bottom approaches. In the former, one uses low-energy observables as boundary conditions for the renormalization 
group equations (RGEs) that run parameters to $M_G$. In this case YU cannot be imposed exactly, but is obtained within a given tolerance. In the second case, instead, 
boundary conditions are specified at $M_G$ (and $\mu, \tan\beta$ at low energy), and in particular {\em exact} YU can be enforced. Thereafter RGE running allows to 
obtain parameters at low energies, whence observables are calculated; the latter can then be fitted to experimental data. 
The two approaches may give sometimes discrepant results. 
In this respect, it is worthwhile to underscore again the strong IR-sensitivity of the YU condition at $M_G$, because of the mentioned threshold corrections. In this 
case, in order to be able to properly address the question of what parameter space is compatible with YU, one should refrain from limiting low-energy input to just 
central values: small variations in the latter mean in general substantial variations in YU. This point has been cleared up in \cite{TobeWells}.

When studying the parameter space of YU, eq. (\ref{ISMH}), in the light of quark masses and FCNCs, an interesting finding is that light (i.e. sub-TeV) SUSY can be made 
compatible with both classes of observables by advocating a $b \to s \gamma$ amplitude $C_7$ simply reversed in sign with respect to the SM one \cite{BDR}. 
Qualitatively, since the dominant contributions, from charginos, interfere destructively with the SM ones, they can be made (at the SUSY-threshold scale) large enough so that, 
at the physical scale of the process, $\mu_b =$O($m_b$), one has $C_7 \simeq -C_7^{\rm SM}$. The branching ratio, going as $|C_7|^2$ will not
be sensitive to the sign flip. This possibility is however disfavored, on a model-independent basis, by data on $B \to X_s \ell^+ \ell^-$, as shown in \cite{GHM}.

The question of the viability of this solution, as well as of exact YU as a whole, has been reappraised in \cite{AABuGuS}. Here, an SO(10) SUSY GUT model proposed by
Dermisek and Raby (DR) \cite{DR}, and featuring Yukawa unification as well as a family symmetry for Yukawa textures, has been reconsidered in a global analysis in the 
light, among the other observables, of all the most precise data on FCNCs in the quark sector. While the model successfully describes EW observables as well as quark 
and lepton masses and mixings \cite{DR,DHR}, ref. \cite{AABuGuS} showed that the simultaneous description of these observables {\em and} all the FCNC processes considered 
is impossible unless the squark masses are pushed well above the limits allowed by naturalness and within reach of the LHC experiments.

\section{A go/no-go global analysis using FCNCs}

\noi The findings of ref. \cite{AABuGuS} prompted two general questions. The first is whether the tension encountered in the simultaneous description of all data, including 
FCNCs, is a general feature of SUSY GUT models with YU and universal sfermion and gaugino mass terms at the GUT scale, thus challenging (barring decoupling in the squark sector) 
the viability of these hypotheses, when considered together. 

The second question concerns a potential remedy to the problem. Namely, one can note that lowering $\tan \beta$ 
alleviates the pressure from $B_s \to \mu^+ \mu^-$, permitting in turn larger Higgs and smaller chargino contributions to $B \to X_s \gamma$, thereby making possible that those 
two contributions indeed cancel to a large extent. Lowering $\tan \beta$ means relaxing $t - b - \tau$ YU to the less restrictive $b - \tau$ unification, as it occurs e.g. in 
SU(5). One should stress that
this solution is non trivial, since $b - \tau$ unification requires $\tan \beta$ either close to unity (which is however excluded by the Higgs mass bound \cite{LEPHiggs}) 
or O(50), because otherwise the predicted bottom quark mass is in general too large \cite{CPW}. Although the case $\tan \beta =$ O(50) can be significantly 
modified by the $\tan \beta$-enhanced threshold corrections to $m_b$ mentioned above, $b - \tau$ unification is difficult to achieve for $\tan\beta \lesssim 35$. Therefore 
$b - \tau$ unification pushes by itself $\tan \beta$ to {\em high} values. Hence the second question: is $\tan \beta$ lower than 50 a viable remedy?

\begin{figure*}[tb]
\begin{center}
\includegraphics[width=0.95 \textwidth]{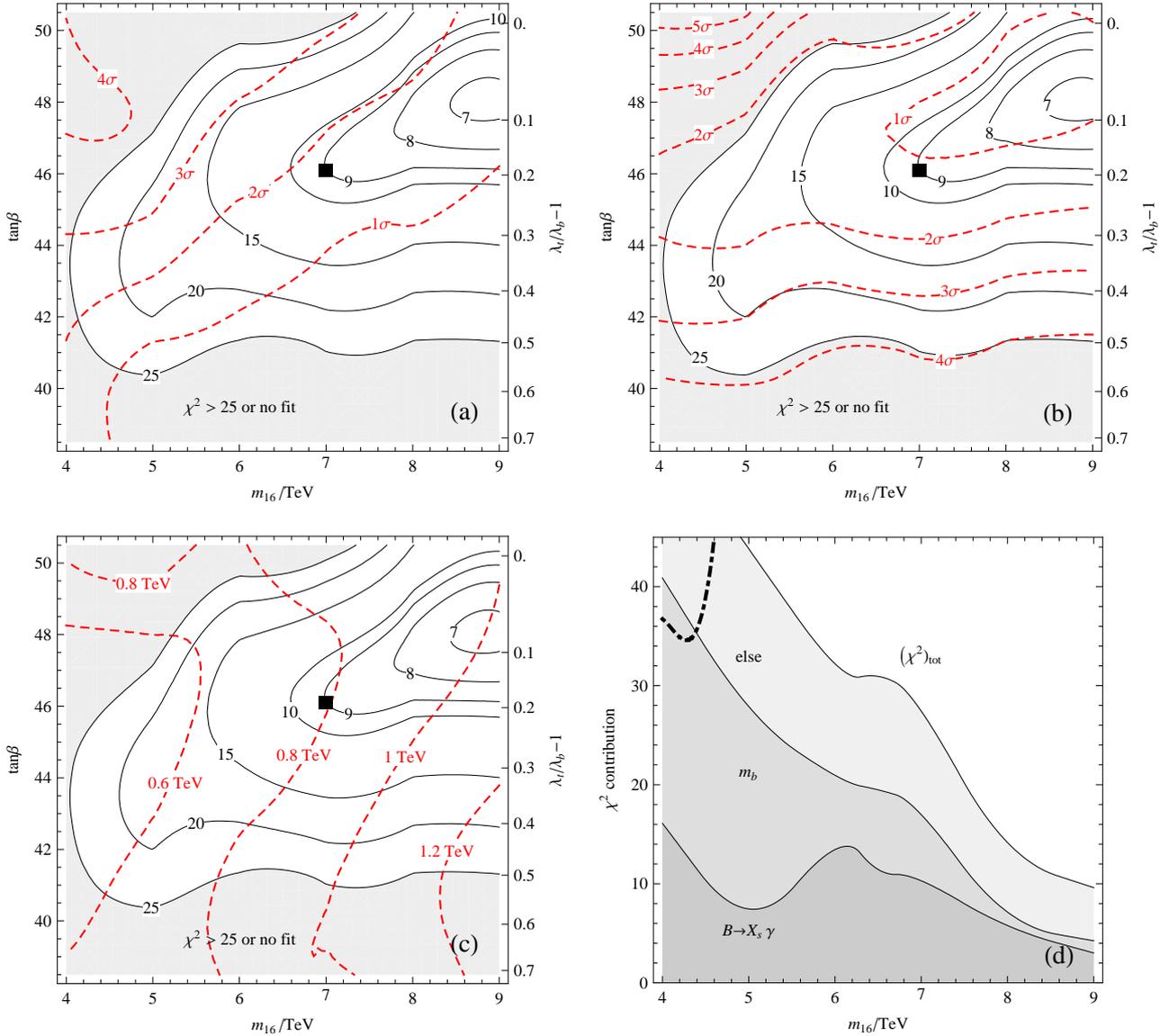}
\caption{Panels (a)-(c): $\chi^2$ contours (solid lines) in the $m_{16}$ vs. $\tan \beta$ plane.
Superimposed as dashed lines are the pulls for BR$(B \to X_s \gamma)$ (panel (a)) and for $m_b$ 
(panel (b)) and the lightest stop mass contours (panel (c)).
Panel (d): $\chi^2$ contributions vs. $m_{16}$ in the special case of exact Yukawa unification.
All the plots assume a SM-like sign for the $b \to s \gamma$ amplitude, except for panel (d), 
where also the total $\chi^2$ for the flipped-sign case is shown as a dot-dashed line. From ref. \cite{AlGuRaS}.}
\label{fig}
\end{center}
\end{figure*}

These questions have been addressed in ref. \cite{AlGuRaS}, where the parameter space of the mentioned class of models has been explored through a $\chi^2$ minimization 
procedure. The adopted strategy, including the considered observables, is in most respects analogous to ref. \cite{AABuGuS}. However, in \cite{AlGuRaS} no assumptions aside 
from YU have been made on Yukawa textures. Results are displayed in the four panels of fig. \ref{fig} and can be summarized as follows: {\bf \em (i.)} panel (d) shows that the 
hypothesis of exact YU is generally challenged, unless the lightest stop is pushed above around 1.1 TeV (see upper part of panel (c) as well), the other squarks being in 
the multi-TeV range. This statement holds under the prior assumption that the sign of the $b \to s \gamma$ amplitude be the same as in the SM. For $m_{16} \lesssim 4.7$ TeV, 
fits prefer the flipped-sign solution, but in this instance, on top of a $\gtrsim 3 \sigma$ discrepancy in $B \to X_s \ell^+ \ell^-$, a true agreement between $b \to s \gamma$ 
and, simultaneously, EW observables and/or $m_b$ is difficult to achieve; {\bf \em (ii.)} when lowering $\tan \beta$, the tension is in fact largely relieved. 
The fit clearly prefers large values of $46 \lesssim \tan \beta \lesssim 48$, as a compromise between FCNCs and $m_b$, that push $\tan \beta$ to respectively lower (see panel (a)) 
and larger values (see panel (b)). The range for $\tan \beta$ corresponds to a moderate breaking of $t - b$ Yukawa unification, in the interval 10-20\%.

In the interesting region, we find the lightest stop mass $\gtrsim 800$ GeV, a light gluino around 400 GeV and lightest Higgs, neutralino and chargino close to the lower bounds. 
This spectrum implies BR$(B_s \to \mu^+ \mu^-)$ in the range 2 to 4$\times 10^{-8}$ and BR$(B \to X_s \gamma)$ robustly around $2.9 \times 10^{-4}$. The requirement of $b -\tau$ 
unification along with the cross fire of the $m_b$ and FCNC constraints are enough to make the above figures basically a firm prediction within the interesting region, 
hence falsifiable at the LHC. See \cite{BaerLHC} for a recent study addressing this point.

\begin{acknowledgments}

\noi It is a pleasure to thank the co-authors of refs. \cite{AABuGuS,AlGuRaS} for the most enjoyable collaboration and for useful comments on the manuscript. 
I also wish to acknowledge the A. von Humboldt Stiftung for kind support. This work has also been supported in part by the Cluster of Excellence ``Origin and 
Structure of the Universe''.

\end{acknowledgments}

\bibliography{ICHEP_proc}

\end{document}